\documentclass[lettersize,journal]{IEEEtran}
\usepackage{amsmath,amsfonts}
\usepackage{algorithmic}
\usepackage{algorithm}
\usepackage{array}
\usepackage[caption=false,font=normalsize,labelfont=sf,textfont=sf]{subfig}
\usepackage{textcomp}
\usepackage{stfloats}
\usepackage{url}
\usepackage{verbatim}
\usepackage{graphicx}
\usepackage{cite}
\usepackage{graphicx}
\usepackage{textcomp}
\usepackage{xcolor}
\usepackage{tikz}
\usetikzlibrary{positioning,arrows.meta}
\usepackage{booktabs}

\begin{document}

\title{Spectral-Leakage-Guided Phase-Noise Compensation for Sub-THz OFDM}

\author{Ahmadreza Afi$^{1, 2}$, Didem Aydogan$^{2}$, Korkut Kaan Tokgöz$^{2}$\\
\footnotesize{$^{1}$Dept. of Electrical Engineering, Amirkabir University of Technology (Tehran Polytechnic), Tehran, Iran}\\
\footnotesize{$^{2}$Faculty of Engineering and Natural Sciences, Electronics Engineering Department, Sabancı University, İstanbul, Türkiye}\\
\scriptsize{email: aaafi@aut.ac.ir, didem.aydogan@sabanciuniv.edu, korkut.tokgoz@sabanciuniv.edu}%
\thanks{The authors thank the Program for Undergraduate Research (PURE) at Sabancı University, where this research began, and the Türkiye Scholarships for funding provided during the internship.}
}
\markboth{Journal of \LaTeX\ Class Files,~Vol.~14, No.~8, August~2021}%
{Shell \MakeLowercase{\textit{et al.}}: A Sample Article Using IEEEtran.cls for IEEE Journals}


\maketitle

\begin{abstract}
Oscillator phase noise is a key impairment in sub-THz OFDM,
where it spreads energy across subcarriers and generates
inter-carrier interference (ICI). Since the resulting leakage
depends on both the oscillator phase noise spectrum and OFDM
numerology, the useful compensation dimension varies with the
operating configuration. This letter develops a leakage-guided
receiver framework that relates these characteristics to the
retained dimension of dominant-component Least Squares (LS)
compensation and its clustered-pilot support. At 130~GHz with 64-QAM, the minimum tested order
achieving $\mathrm{BER}\leq10^{-6}$ decreases from 9 at
240~kHz to 3 at 960--1920~kHz for Hexa-X Model~2, while the first 3GPP crossing at 1920~kHz occurs with
$N_p=13$ at 38~dB. Order adaptation
reduces Stage-2 arithmetic by 66\%, while four-cluster pilots
improve 32-dB goodput by 35\% over comb allocation.
\end{abstract}


\section{Introduction}
\label{sec:introduction}

The move toward 6G is expected to push wireless systems into sub-Terahertz (sub-THz) frequency ranges to support both terabit-per-second communications \cite{usitalo20216g,HexaX2021} and integrated sensing and communication (ISAC) \cite{aydogan2026impacts}.
While these bands offer substantial capacity, transceiver impairments become increasingly significant as the carrier frequency rises. Among them, oscillator phase noise becomes particularly
severe, particularly for
multicarrier waveforms such as Orthogonal Frequency Division
Multiplexing (OFDM) \cite{
aydogan2025comparative, armada2001understanding, chung2022phase}.

In OFDM, phase noise acts as a multiplicative time-domain
impairment: the zero-order spectral component produces the Common
Phase Error (CPE), observed as a common phase rotation across the
subcarriers of an OFDM symbol, while the nonzero components couple
energy between subcarriers, producing spectral leakage and
Inter-Carrier Interference (ICI)
\cite{Petrovic2004,chung2022phase}. The extent and distribution
of this leakage depend jointly on the oscillator phase noise spectrum
and the OFDM numerology, particularly the subcarrier spacing.
Consequently, the number and relative strength of spectrally
significant phase-noise components can vary considerably across
operating conditions, leading to different compensation requirements
\cite{aydogan2025comparative,aydogan2024homodyne}.

A large body of work has addressed modeling and mitigation
of phase-noise in OFDM. Classical studies established the CPE
and ICI behavior caused by phase-noise
\cite{armada2001understanding,Petrovic2004}, followed by
pilot-assisted and code-aided approaches for phase tracking and
estimation \cite{Petrovic_kalman,sibel2018pilot}. More recent methods jointly
estimate and compensate phase-noise by exploiting its
frequency-domain structure or temporal correlation
\cite{chung2022phase,linsalata2022joint}. At sub-THz frequencies, the choice of phase-noise model for accurate performance estimation becomes
increasingly important: distinct phase-noise models, such as Hexa-X model 2 and
3GPP TR~38.803 \cite{HexaX2021, 3gpp_tr38803},
 produce different spectral-leakage patterns and can
therefore lead to different compensation requirements \cite{aydogan2025comparative}.

These differences in spectral leakage have direct consequences
for receiver design. Frequency-domain phase-noise compensation
requires selecting how many spectral components should be
estimated and how much pilot support should be allocated to
their estimation \cite{chung2022phase}. Retaining too few
components leaves residual ICI, whereas retaining more than
needed increases the estimation dimension, pilot overhead, and
processing cost
\cite{armada2001understanding,chung2022phase}. The selected
order therefore affects not only the residual interference, but
also the amount of pilot support required for reliable
estimation. Since the relevant leakage support changes with both
the phase-noise spectrum and subcarrier spacing
\cite{aydogan2025comparative}, the compensation order and its
pilot support should be selected jointly according to the
underlying leakage behavior.

To overcome this critical bottleneck, we introduce a highly efficient, leakage-guided phase-noise compensation framework tailored for sub-THz OFDM at 130~GHz. We characterize hardware-dependent spectral leakage using the Hexa-X Model~2 phase-noise model and compare it with the 3GPP TR~38.803 reference. By accounting for how spectral leakage varies with OFDM
numerology, the framework guides the retained compensation
dimension $N_p$ and the associated pilot-observation support
required for LS estimation. The receiver then uses a scalar
Kalman filter to track the symbol-wise CPE, followed by LS
estimation of the retained dominant phase-noise coefficients
from distributed clustered pilots. Under the evaluated conditions, the leakage-guided framework
achieves $\mathrm{BER}\leq10^{-6}$ using only one third of the
retained compensation dimension, reducing Stage-2 arithmetic
complexity by 66\%. The four-cluster pilot design further
achieves a 35\% goodput gain at 32~dB over the comb allocation,
adding a throughput benefit alongside the complexity reduction.

This letter is organized as follows. Section~\ref{sec:system_model}
introduces the system model and leakage analysis,
Section~\ref{sec:proposed_method} presents the compensation and pilot design, Section~\ref{sec:results} discusses the numerical
results, and Section~\ref{sec:clc} concludes the paper.
\section{System Model and Phase-Noise Leakage}
\label{sec:system_model}
\subsection{OFDM-Based Waveform Model}
\label{subsec:ofdm_model}

We consider an OFDM system with $N$ subcarriers, where normalized
unit-power $M$-QAM symbols $X[k]$ are mapped onto the frequency-domain
resource grid. After IFFT processing and cyclic-prefix insertion, the
resulting time-domain OFDM signal $x[n]$ is impaired by oscillator
phase-noise $\phi[n]$ and additive white Gaussian noise (AWGN) $w[n]$. After cyclic-prefix removal, the received baseband samples are modeled as:
\begin{equation}
    y[n] = x[n] e^{j\phi[n]} + w[n].
    \label{eq:rx_time}
\end{equation}

Rather than invoking a small-angle approximation, the effect of phase noise is characterized directly through the discrete Fourier transform
of its complex envelope. The corresponding phase-noise spectral
coefficients are given as in \cite{Petrovic2004,Petrovic_kalman}:
\begin{equation}
    J_{\ell}
    =
    \frac{1}{N}
    \sum_{n=0}^{N-1}
    e^{j\phi[n]}
    e^{-j2\pi \ell n/N},
    \qquad
    \ell=0,\ldots,N-1 .
    \label{eq:pn_coefficients}
\end{equation}

The received signal on the $k$th subcarrier can then be expressed as:
\begin{equation}
    Y[k]
    =
    J_{0}X[k]
    +
    \sum_{\substack{i=0\\i\neq k}}^{N-1}
    J_{k-i}X[i]
    +
    W[k],
    \label{eq:rx_frequency}
\end{equation}
where all spectral indices are interpreted modulo $N$. Equation
\eqref{eq:rx_frequency} separates the phase-noise impairment into two
components. The coefficient $J_{0}$ represents the common phase error
(CPE), which produces a common rotation and amplitude perturbation
across the OFDM symbol, whereas the coefficients $J_{\ell}$,
$\ell\neq0$, generate ICI through spectral
leakage among subcarriers and $W[k]$ is the AWGN. 
This frequency-domain representation forms
the basis for the leakage analysis and compensation-order selection
developed in the following subsections.
\subsection{Phase-Noise Modeling and Spectral Leakage}
\label{subsec:pn_profiles}
\subsubsection{Phase-Noise Models}
To capture hardware-dependent phase-noise behavior at sub-THz
frequencies, we consider the Hexa-X Model~2 at
$f_c=130$~GHz as a baseline and the 3GPP TR~38.803 model
as a reference comparison \cite{HexaX2021, 3gpp_tr38803}.
The two models exhibit distinct leakage characteristics.
Hexa-X shows rapidly decaying close-in leakage, concentrating
most ICI energy around neighboring subcarriers, whereas the
3GPP model exhibits a broader and non-monotonic spectrum
that spreads interference over a larger range of offsets.

For a given subcarrier spacing $\Delta f$, coefficient $J_\ell$
samples the oscillator phase noise spectrum at approximately
\begin{equation}
    f_\ell=|\ell|\Delta f .
\end{equation}
Thus, interference spread depends jointly on the phase-noise model and OFDM numerology. Smaller $\Delta f$ captures more energetic leakage components, while wider spacing concentrates rapidly decaying leakage into fewer terms. Consequently, the required compensation order must carefully balance residual ICI against estimation overhead. While rapidly decaying phase-noise models need fewer components at wider SCS, broader spectra (e.g., 3GPP) spread leakage over more offsets and exhibit less SCS dependence.

\subsubsection{Residual Leakage and Compensation-Order Guidance}
For a retained compensation order $N_p=2\gamma+1$, where
$\gamma\geq0$ is the truncation radius, the receiver models the
phase-noise coefficients with indices
$\ell=-\gamma,\ldots,\gamma$. Define the average power of the
$\ell$th phase-noise coefficient as:
\begin{equation}
    E_\ell = \mathbb{E}\left\{|J_\ell|^2\right\},
\end{equation}
where $J_\ell$ is given in \eqref{eq:pn_coefficients}. Since the
time-domain phase-noise envelope has unit magnitude,
Parseval's relation gives:
\begin{equation}
    \sum_{\ell=0}^{N-1} E_\ell = 1.
\end{equation}
Accordingly, the residual leakage outside the retained spectral
support is
\begin{equation}
    \mathcal{L}_{\mathrm{res}}(\gamma)
    =
    1-
    \sum_{\ell=-\gamma}^{\gamma}E_\ell,
    \label{eq:residual_leakage}
\end{equation}
where the coefficient indices are interpreted modulo $N$.

Increasing $\gamma$ by one adds the two spectral components
at offsets $\pm\gamma$. Hence, for $\gamma\geq1$, the
incremental reduction in residual leakage is:
\begin{equation}
    \Delta\mathcal{L}_{\mathrm{res}}(\gamma)
    =
    \mathcal{L}_{\mathrm{res}}(\gamma-1)
    -
    \mathcal{L}_{\mathrm{res}}(\gamma)
    =
    E_{\gamma}+E_{-\gamma}.
    \label{eq:incremental_leakage}
\end{equation}
Thus, $\Delta\mathcal{L}_{\mathrm{res}}(\gamma)$ quantifies
the additional phase-noise power captured when the retained
order is increased from $2\gamma-1$ to $2\gamma+1$.

For a slowly decaying leakage spectrum, these incremental
contributions remain significant over a wider range of $\gamma$,
indicating greater sensitivity to compensation order. For a
rapidly decaying spectrum,
$\Delta\mathcal{L}_{\mathrm{res}}(\gamma)$ becomes small after
only a few neighboring components are retained, yielding
diminishing reductions in residual leakage. The leakage behavior therefore indicates the useful compensation
dimension. Since performance also depends on estimation error,
pilot geometry, SNR, and modulation, the operational order is
determined from the BER results in Section~\ref{sec:results}.
We report the minimum tested $N_p$ achieving the target BER for
each phase-noise spectrum and OFDM numerology.
\section{Leakage-Aware Compensation and Pilot Design}
\label{sec:proposed_method}
The receiver uses two-stage phase-noise compensation based on the conventional CPE/ICI decomposition \cite{armada2001understanding,chung2022phase}. Stage~1 tracks and corrects the pilot-based symbol-wise CPE using a scalar Kalman filter. Stage~2 estimates dominant residual phase-noise
coefficients using pilot-assisted LS. The retained order sets the estimator dimension, while the pilot allocation determines the observation support and overhead.
\subsection{Clustered Pilot Allocation}
\label{subsec:pilot_design}

\begin{figure}[t]
    \centering
    \includegraphics[width=1\columnwidth]{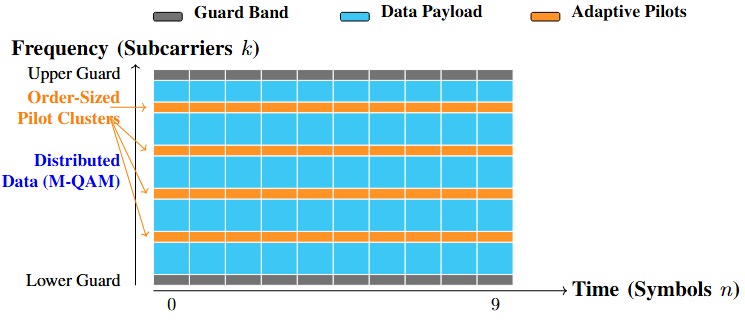}
    \caption{Distributed clustered-pilot allocation used for
phase-noise estimation.}
    \label{fig:cluster_allocation}
\end{figure}

The receiver uses distributed clusters of contiguous pilots to
provide known neighboring symbols for phase-noise estimation.
Compared with a comb allocation, clustering reduces contamination
of the selected pilot observations by adjacent unknown data
symbols. For $K$ clusters of width $C$, the total pilot overhead is
$N_{\mathrm{pil}}=KC$. With retained order
$N_p=2\gamma+1$, excluding $\gamma$ observations at each
cluster edge leaves:
\begin{equation}
    N_{\mathrm{obs}}=K(C-2\gamma)
\end{equation}
protected observations.
Let $\mathbf{A}\in\mathbb{C}^{N_{\mathrm{obs}}\times N_p}$
denote the LS observation matrix formed from the corresponding
known pilot symbols. The evaluated clustered geometries are chosen to satisfy:
\begin{equation}
    N_{\mathrm{obs}}\geq N_p,
    \qquad
    \operatorname{rank}(\mathbf{A})=N_p,
\end{equation}
so that the retained phase-noise coefficients are identifiable.
Figure~\ref{fig:cluster_allocation} illustrates the distributed
resource-grid arrangement. Different cluster counts and widths are
then compared to characterize the trade-off between estimation
reliability and pilot overhead.
\subsection{Two-Stage Phase-Noise Compensation}
\label{subsec:two_stage}
\subsubsection{Stage 1: Kalman-Based CPE Tracking}
Let $z_m$ denote the phase observation obtained from
the pilot subcarriers of the $m$th OFDM symbol.
A scalar Kalman filter tracks this observation using
a random-walk state model with process-noise variance
$Q$ obtained as in \cite{Petrovic_kalman}. The measurement variance is set according to
the SNR-based approximation
$R=(2\rho N_{\mathrm{pil}})^{-1}$, where $\rho$
denotes the linear SNR and $N_{\mathrm{pil}}$
is the number of pilot subcarriers.

With the predicted phase
$\widehat{\theta}_{m|m-1}
=\widehat{\theta}_{m-1|m-1}$, the filter updates are
\begin{equation}
    \begin{aligned}
        P_{m|m-1}
        &=P_{m-1|m-1}+Q,\\
        K_m
        &=\frac{P_{m|m-1}}{P_{m|m-1}+R},\\
        \nu_m
        &=\arg\left\{
        e^{j(z_m-\widehat{\theta}_{m|m-1})}
        \right\},\\
        \widehat{\theta}_{m|m}
        &=\widehat{\theta}_{m|m-1}+K_m\nu_m,\\
        P_{m|m}
        &=(1-K_m)P_{m|m-1},
    \end{aligned}
    \label{eq:kalman_recursion}
\end{equation}
where $P_{m|m-1}$ and $P_{m|m}$ are the predicted
and updated state-error variances, respectively.
The wrapped innovation $\nu_m$ accounts for the
periodicity of phase.

The estimated phase is applied uniformly across
the received OFDM symbol:
\begin{equation}
    y'_m[n]
    =
    y_m[n]e^{-j\widehat{\theta}_{m|m}}.
    \label{eq:cpe_correction}
\end{equation}
The corrected samples are subsequently transformed
to the frequency domain to obtain $Y'_m[k]$.

\subsubsection{Stage 2: Dominant-Component LS Compensation}
The phase-noise spectrum is approximated by retaining
$N_p=(2\gamma+1)$ central components, corresponding to
offsets $\ell=-\gamma,\ldots,\gamma$.
Let $\mathbf{p}_{\gamma,m}$ denote the retained
coefficient vector after the first-stage correction,
including its central component.

The known pilot symbols are used to construct
the frequency-domain observation matrix
$\mathbf{A}_{\gamma,m}
=\mathbf{X}^{p}_{f,\gamma,m}
\in\mathbb{C}^{N_{\mathrm{obs}}\times N_p}$.
For a full-column-rank matrix, the retained
phase-noise coefficients are estimated as:
\begin{equation}
    \widehat{\mathbf{p}}_{\gamma,m}
    =
    \left(
    \mathbf{A}_{\gamma,m}^{H}
    \mathbf{A}_{\gamma,m}
    \right)^{-1}
    \mathbf{A}_{\gamma,m}^{H}
    \mathbf{y}'_{\mathrm{sub},m},
    \label{eq:ls_estimator}
\end{equation}
where $\mathbf{y}'_{\mathrm{sub},m}$ contains the
selected pilot observations from the corrected
frequency-domain signal.
The estimated coefficients are embedded at the retained
spectral offsets and transformed back to the time domain.
Using the DFT convention in \eqref{eq:pn_coefficients}, the reconstructed
phase-noise envelope is:
\begin{equation}
    \hat{q}_m[n]
    =
    \sum_{\ell=-\gamma}^{\gamma}
    \hat{p}_{\ell,m}
    e^{j2\pi \ell n/N}.
    \label{eq:pn_reconstruction}
\end{equation}
Since the physical phase-noise process has unit magnitude, only
the phase of this reconstruction is retained. The Stage-2
compensated samples are therefore:
\begin{equation}
    \tilde{y}_m[n]
    =
    y'_m[n]\,
    e^{-j\arg\{\hat{q}_m[n]\}},
    \label{eq:stage2_compensation}
\end{equation}
before the final FFT and data detection.
The designation KF+LS refers to this processing
sequence. Comparisons against CPE-only correction
evaluate the effect of adding spectral phase-noise
compensation, rather than establishing an independent
benefit of Kalman preprocessing over matched LS-only
processing.
\subsection{Computational Complexity}
\label{subsec:complexity}

For a fixed pilot geometry, the LS pseudoinverse
$\mathbf{A}^{\dagger}$ can be precomputed for each retained
order $N_p$ and reused across OFDM symbols. The dominant
online order-dependent cost therefore arises from the LS
projection and reconstruction of the retained phase-noise
components, requiring approximately
\begin{equation}
    C_{\mathrm{S2}}
    \approx
    N_p\left(N_{\mathrm{obs}}+N\right)
\end{equation}
complex multiplications per OFDM symbol. The KF, FFT, and
other operations common to all retained orders are excluded
from this order-dependent comparison.

\section{Results and Discussion}
\label{sec:results}
The receiver is evaluated for a 130-GHz OFDM link with
$N=1024$ subcarriers, a cyclic prefix of $N_{\mathrm{CP}}=64$
samples, and 64-QAM modulation. The considered subcarrier
spacings are
$\Delta f\in\{240,480,960,1920\}$~kHz, with the SNR swept
from 0 to 40~dB. Unless otherwise stated, the BER evaluations use $K=4$ distributed pilot clusters with $C=49$ pilots per cluster. Each Monte Carlo realization contains 20
OFDM symbols with phase noise evolving continuously across
the frame. Hexa-X Model~2 is considered as the primary phase-noise model to evaluate the proposed framework and the 3GPP TR~38.803 model is used as
a cross-model comparison. Each operating point uses at
least 100 independent realizations, with high-SNR cases
extended to approximately $2\times10^7$ detected bits.
\subsubsection{Spectral Leakage and Compensation-Order Sensitivity}
\label{subsec:leakage_results}

\begin{figure}[t]
    \centering
    \includegraphics[width=1\columnwidth]{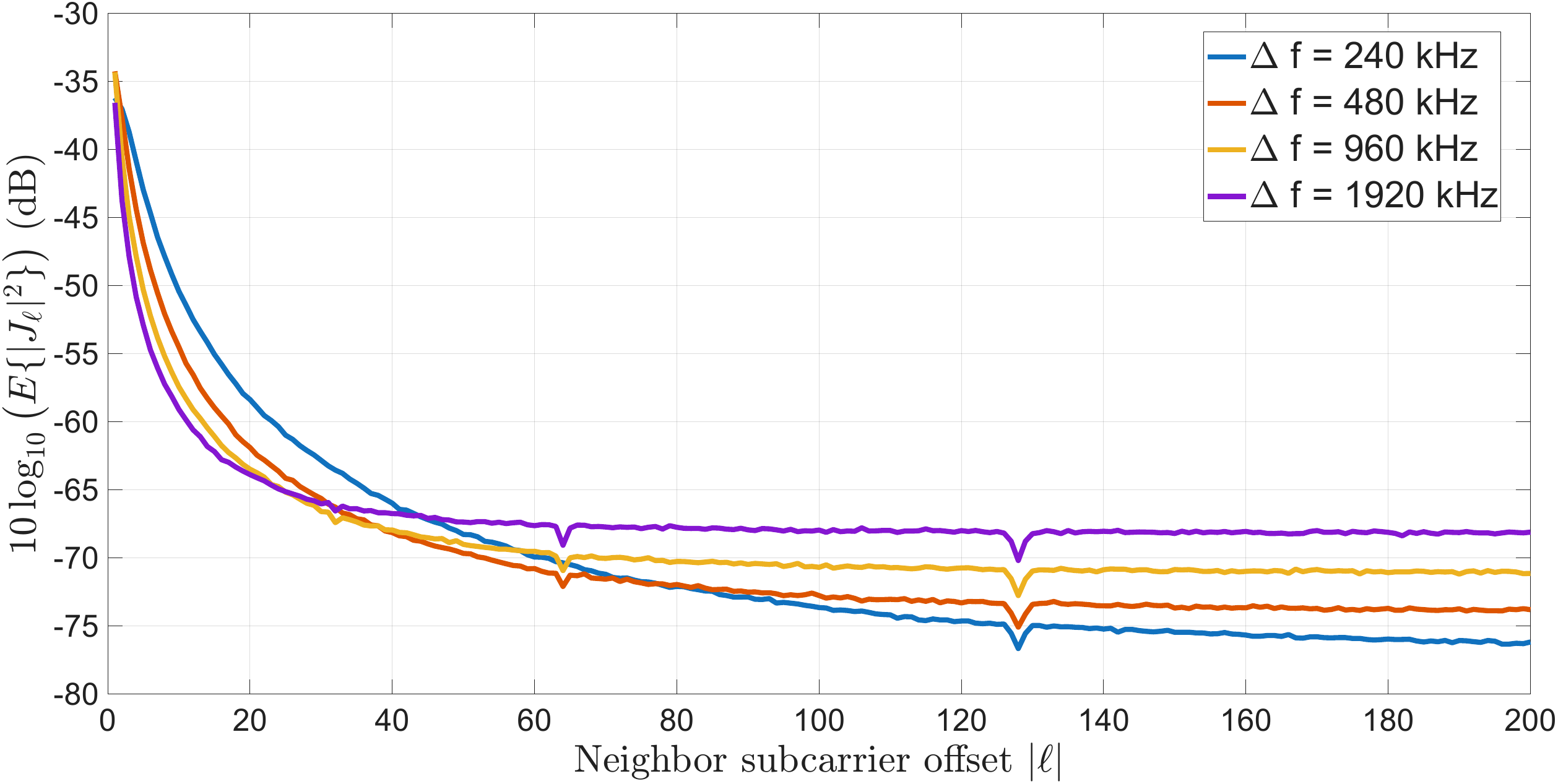}
    \caption{Far-neighbor phase-noise leakage at 130~GHz for the Hexa-X model.}
\label{fig:long_tail_hexax}
\end{figure}

\begin{figure}[h!]
    \centering
    \includegraphics[width=1\columnwidth]{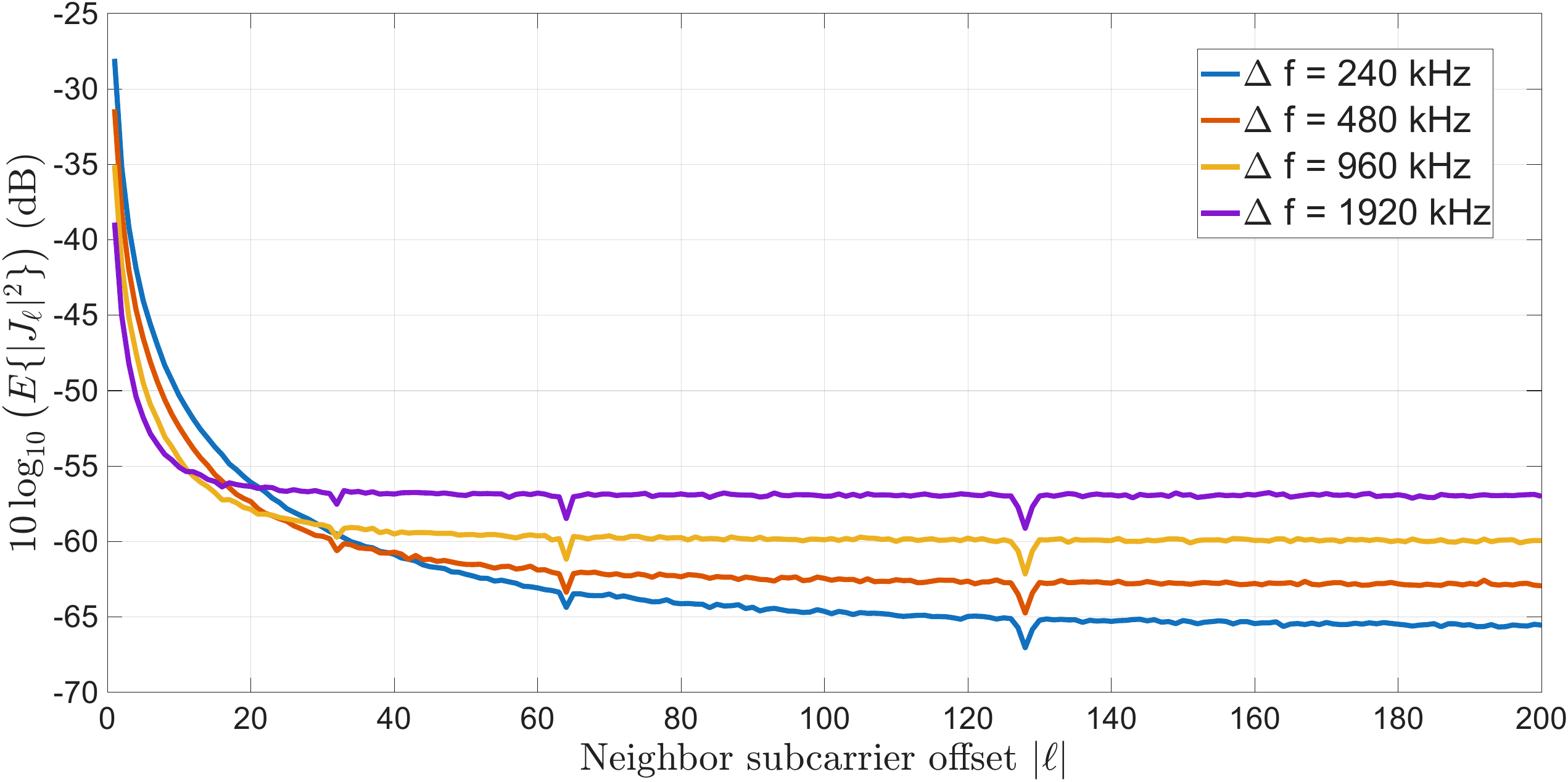}
    \caption{Far-neighbor phase-noise leakage at 130~GHz for the 3GPP model.}
    \label{fig:long_tail_3gpp}
\end{figure}

Figures~\ref{fig:long_tail_hexax} and
\ref{fig:long_tail_3gpp} show the average spectral leakage
$E_\ell=\mathbb{E}\{|J_\ell|^2\}$ at 130~GHz for the considered
subcarrier spacings. The Hexa-X model exhibits rapidly decaying
leakage with increasing subcarrier offset, whereas the 3GPP
model retains a broader and non-monotonic spectral tail.
Consequently, changing $\Delta f$ changes the physical offset
sampled by each coefficient $J_\ell$, so the leakage support
depends on both the oscillator phase noise spectrum and OFDM
numerology. Broader, non-monotonic models can remain significant
over more spectral offsets and therefore exhibit a different
dependence on subcarrier spacing.
\begin{table}[b]
\centering
\caption{Minimum tested compensation order for
$\mathrm{BER}\leq10^{-6}$ at 130~GHz.}
\label{tab:order_scs}
\begin{tabular}{c c c c}
\hline
PN Model & $\Delta f$ & SNR & Min. $N_p$ \\
\hline
Hexa-X & 240~kHz  & 28~dB & 9 \\
Hexa-X & 480~kHz  & 28~dB & 5 \\
Hexa-X & 960~kHz  & 28~dB & 3 \\
Hexa-X & 1920~kHz & 28~dB & 3 \\
3GPP   & 1920~kHz & 38~dB & 13 \\
\hline
\end{tabular}
\end{table}

The receiver-level consequence is summarized in
Table~\ref{tab:order_scs}, which reports the minimum tested
$N_p$ achieving $\mathrm{BER}\leq10^{-6}$ at the earliest
simulated SNR for each configuration. For Hexa-X, this occurs at
28~dB, with the required order decreasing from $N_p=9$ at
240~kHz to $N_p=5$ at 480~kHz and $N_p=3$ at both
960 and 1920~kHz. The 3GPP model provides a cross-model
comparison at 1920~kHz, where the first crossing of the same BER
target occurs at 38~dB with $N_p=13$. The larger required order
is consistent with the broader leakage observed in
Fig.~\ref{fig:long_tail_3gpp}.

These results show that a single fixed compensation order is not
suitable across different phase-noise models and numerologies.
The SCS dependence follows from mapping the continuous
phase-noise spectrum onto discrete ICI coefficients: at smaller
$\Delta f$, more coefficients fall within the dominant portion of
the Hexa-X spectrum, whereas larger $\Delta f$ places the sampled
offsets farther apart and concentrates significant leakage into
fewer neighboring terms. Once these dominant components are
retained, further increases in $N_p$ provide diminishing benefit,
explaining the smaller useful order at wider subcarrier spacings.

\subsubsection{BER Performance}
\label{subsec:ber_results}
\begin{figure*}[t]
    \centering \includegraphics[width=1\textwidth]{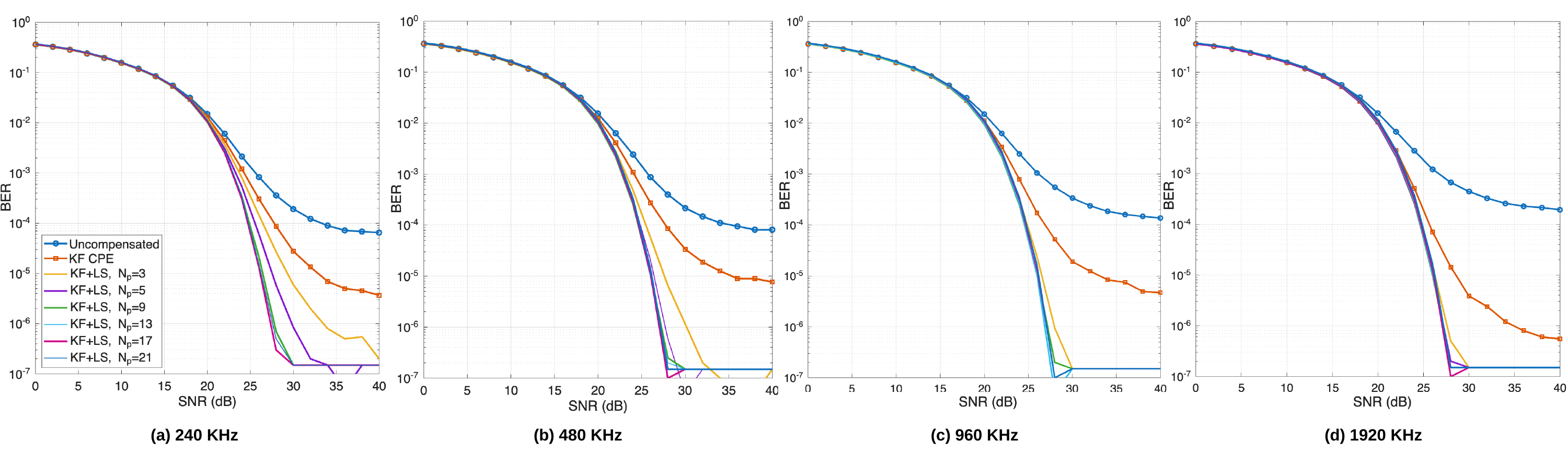}
    \caption{BER under Hexa-X model 2 phase-noise at 130~GHz with 64-QAM for
    different subcarrier spacings and compensation orders.}
\label{fig:hexax_results}
\end{figure*}

Figure~\ref{fig:hexax_results} shows the resulting BER under
Hexa-X Model~2 for
$N_p\in\{3,5,9,13,17,21\}$. Zero-error operating points are
shown using the corresponding 95\% upper confidence bound.
The useful order decreases as the subcarrier spacing increases.
At 240~kHz and 28~dB, the BER improves from
$2.70\times10^{-5}$ for $N_p=3$ to
$6.98\times10^{-7}$ for $N_p=9$ as given in Table~\ref{tab:order_scs}.
Once the
dominant leakage components are captured, increasing $N_p$
provides diminishing benefit and can slightly degrade BER
because additional coefficients must be estimated from noisy
pilot observations.
The broader 3GPP model provides a contrasting case. At
$\Delta f=1920$~kHz, the first $10^{-6}$ crossing occurs at
38~dB with $N_p=13$, where the BER is
$8.98\times10^{-7}$, while $N_p=3$, 5, and 9 remain above
the target. This confirms that the useful compensation
dimension depends jointly on phase-noise leakage and OFDM
numerology. Finally, the CPE-only baseline exhibits a residual
high-SNR impairment, whereas adding dominant-component LS
compensation enables substantially lower BER. This comparison
demonstrates the benefit of spectral compensation beyond CPE
correction, without isolating an independent gain from the
Kalman preprocessing itself.

Using the BER-validated orders, the Stage-2 cost decreases from
$10.7\times10^3$ to $3.6\times10^3$ CMs per OFDM symbol,
corresponding to a 66\% reduction in arithmetic complexity and
lower implementation burden. Prior FPGA implementations indicate that lower
hardware-resource demand can accompany lower power
consumption~\cite{deCampos2021}, highlighting the potential of the reduced Stage-2 complexity for lower energy consumption.

\subsubsection{Pilot Overhead and Goodput}
\label{subsec:goodput_results}

The pilot-topology trade-off is evaluated at
$\Delta f=240$~kHz with $N_p=9$, as identified by the BER
analysis. For $N_{\mathrm{data}}$ data-bearing subcarriers, the
raw rate including cyclic-prefix overhead is:
\begin{equation}
    R_{\mathrm{raw}}
    =
    N_{\mathrm{data}}\log_2(M)\Delta f
    \frac{N}{N+N_{\mathrm{CP}}},
    \label{eq:raw_rate}
\end{equation}
For an uncoded packet of $L=12\,288$ bits, the BER-derived
goodput is approximated as in~\cite{Liu2026SparkLink}:
\begin{equation}
    G=R_{\mathrm{raw}}(1-\mathrm{BER})^{L},
    \label{eq:goodput}
\end{equation}
\begin{figure}[ht]
    \centering
    \includegraphics[width=0.8\columnwidth]
    {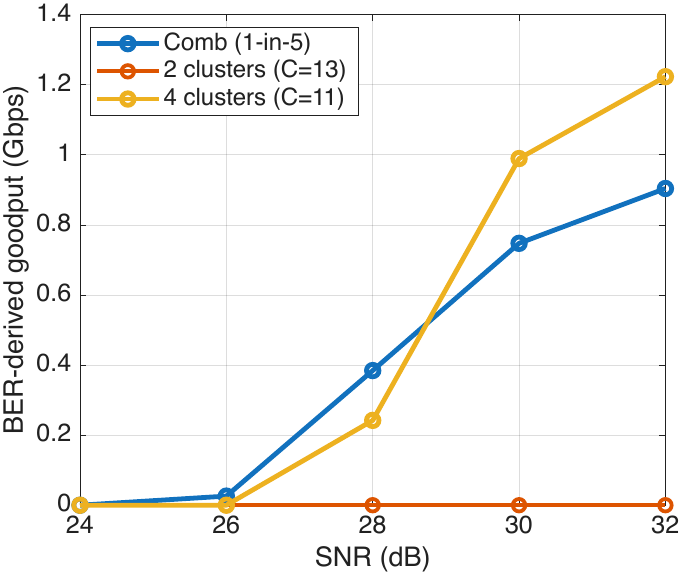}
    \caption{BER-derived goodput for the evaluated pilot
    topologies at 240~kHz with $N_p=9$.}
    \label{fig:goodput_topology}
\end{figure}
assuming independent bit errors; thus, $G$ is used as a
throughput proxy. The comb, two-cluster, and four-cluster
configurations use 195, 26, and 44 pilots, yielding raw rates of
1.054, 1.283, and 1.259~Gbps, respectively.
Figure~\ref{fig:goodput_topology} shows that lower pilot overhead
does not necessarily provide higher goodput. At 28~dB, the comb
allocation achieves 0.384~Gbps versus 0.242~Gbps for the
four-cluster case, whereas at 32~dB the four-cluster design
reaches 1.223~Gbps versus 0.904~Gbps, a gain of about 35\%.
The two-cluster case remains reliability-limited despite its
higher raw-rate ceiling, highlighting the trade-off between pilot
overhead and phase-noise estimation reliability.
\section{Conclusion}\label{sec:clc} 
This letter developed a leakage-guided phase-noise compensation
framework that relates oscillator-dependent spectral leakage and
OFDM numerology to the useful compensation order. For Hexa-X
Model~2, the target $\mathrm{BER}\leq10^{-6}$ is achieved at
wider subcarrier spacings using only one third of the retained
compensation dimension, reducing Stage-2 arithmetic by 66\%.
The broader 3GPP profile, in contrast, requires a larger retained
order. The four-cluster pilot design further provides a 35\%
goodput gain over comb allocation. These results show that
adapting the compensation order and pilot support to
phase-noise-induced spectral leakage can substantially reduce
receiver complexity while improving throughput.
\bibliographystyle{IEEEtran}
\bibliography{references}

@inproceedings{Petrovic2004,
  author={Petrovic, D. and Rave, W. and Fettweis, G.},
  booktitle={IEEE International Conference on Communications, 2005. ICC 2005. 2005}, 
  title={Properties of the intercarrier interference due to phase noise in OFDM}, 
  year={2005},
  volume={4},
  number={},
  pages={2605-2610 Vol. 4},
  doi={10.1109/ICC.2005.1494821}}

@INPROCEEDINGS{Petrovic_kalman,
  author={Petrovic, D. and Rave, W. and Fettweis, G.},
  booktitle={2004 IEEE 15th International Symposium on Personal, Indoor and Mobile Radio Communications (IEEE Cat. No.04TH8754)}, 
  title={Common phase error due to phase noise in OFDM-estimation and suppression}, 
  year={2004},
  volume={3},
  number={},
  pages={1901-1905 Vol.3},
  doi={10.1109/PIMRC.2004.1368329}}

@misc{HexaX2021,
  author      = {{Hexa-X}},
  title       = {Deliverable D2.3: Radio Models and Enabling
                 Techniques Towards Ultra-High Data Rate Links
                 and Capacity in 6G},
  institution = {Hexa-X Project},
  year        = {2023},
  month       = apr,
  note        = {Version 1.0}
}

@article{usitalo20216g,
  author={Uusitalo, Mikko A. and Rugeland, Patrik and Boldi, Mauro Renato and Strinati, Emilio Calvanese and Demestichas, Panagiotis and Ericson, Mårten and Fettweis, Gerhard P. and Filippou, Miltiadis C. and Gati, Azeddine and Hamon, Marie-Helene and Hoffmann, Marco and Latva-Aho, Matti and Pärssinen, Aarno and Richerzhagen, Björn and Schotten, Hans and Svensson, Tommy and Wikström, Gustav and Wymeersch, Henk and Ziegler, Volker and Zou, Yaning},
  journal={IEEE Access}, 
  title={6G Vision, Value, Use Cases and Technologies From European 6G Flagship Project Hexa-X}, 
  year={2021},
  volume={9},
  number={},
  pages={160004-160020},
  doi={10.1109/ACCESS.2021.3130030}}

@article{chung2022phase,
  author={Chung, Minkeun and Liu, Liang and Edfors, Ove},
  journal={IEEE Transactions on Wireless Communications}, 
  title={Phase-Noise Compensation for OFDM Systems Exploiting Coherence Bandwidth: Modeling, Algorithms, and Analysis}, 
  year={2022},
  volume={21},
  number={5},
  pages={3040-3056},
  doi={10.1109/TWC.2021.3117782}}

@inproceedings{aydogan2025comparative,
  author={Aydoğan, Didem and Chraiti, Mohaned and Tokgöz, Korkut Kaan},
  booktitle={2025 Joint European Conference on Networks and Communications \& 6G Summit (EuCNC/6G Summit)}, 
  title={A Comparative Study on Phase Noise Model in Ultra-High Data-Rate Sub-THz Communications}, 
  year={2025},
  volume={},
  number={},
  pages={679-684},
  doi={10.1109/EuCNC/6GSummit63408.2025.11036859}}

@misc{aydogan2024homodyne,
      title={Homodyne vs. Heterodyne Architectures in Sub-THz Transceivers: A Phase Noise Perspective}, 
      author={Didem Aydogan and Korkut Kaan Tokgöz},
      year={2026},
      eprint={2603.21384},
      archivePrefix={arXiv},
      primaryClass={eess.SP},
      url={https://arxiv.org/abs/2603.21384}, 
}

@inproceedings{sibel2018pilot,
  author={Sibel, Jean-Christophe},
  booktitle={2018 25th International Conference on Telecommunications (ICT)}, 
  title={Pilot-Based Phase Noise Tracking for Uplink DFT-s-OFDM in 5G}, 
  year={2018},
  volume={},
  number={},
  pages={52-56},
  doi={10.1109/ICT.2018.8464891}}

@article{armada2001understanding,
  author={Garcia Armada, A.},
  journal={IEEE Transactions on Broadcasting}, 
  title={Understanding the effects of phase noise in orthogonal frequency division multiplexing (OFDM)}, 
  year={2001},
  volume={47},
  number={2},
  pages={153-159},
  doi={10.1109/11.948268}}

@misc{linsalata2022joint,
  author={Linsalata, Francesco and Ksairi, Nassar},
  booktitle={ICASSP 2023 - 2023 IEEE International Conference on Acoustics, Speech and Signal Processing (ICASSP)}, 
  title={On the Joint Estimation of Phase Noise and time-Varying Channels for OFDM under High-Mobility Conditions}, 
  year={2023},
  volume={},
  number={},
  pages={1-5},
  doi={10.1109/ICASSP49357.2023.10094589}}

@techreport{3gpp_tr38803,
  author      = {{3rd Generation Partnership Project (3GPP)}},
  title       = {Study on new radio access technology: Radio Frequency ({RF}) and co-existence aspects},
  type        = {Technical Report (TR)},
  number      = {TR 38.803},
  institution = {3GPP},
  version     = {14.5.0},
  year        = {2017},
  url         = {https://www.3gpp.org/dynareport/38803.htm}
}

@article{deCampos2021,
author = {Fabrício P. V. de Campos  and Lucas M. Sirimarco  and Marcello L. R. de Campos  and Moisés V. Ribeiro },
title = {The implementation of the clustered‐OFDM‐based transceiver on an FPGA device: A comprehensive comparison},
journal = {IET Communications},
volume = {15},
issue = {6},
pages = {824-839},
year = {2021},
doi = {10.1049/cmu2.12124},
URL = {https://digital-library.theiet.org/doi/abs/10.1049/cmu2.12124},
eprint = {https://digital-library.theiet.org/doi/pdf/10.1049/cmu2.12124}
}

@article{Liu2026SparkLink,
author = {Liu, Boya and Ren, Zhenyu and Zhang, Yanbo and Li, Mo and Liang, Jiaxin and He, Bin and Li, Jie and Wu, Hao and Zhou, Jingbin},
title = {Unveiling SparkLink Low Energy: A Comparative Measurement Study},
year = {2026},
issue_date = {February 2026},
publisher = {Association for Computing Machinery},
address = {New York, NY, USA},
volume = {7},
number = {1},
url = {https://doi.org/10.1145/3776559},
doi = {10.1145/3776559},
journal = {ACM Trans. Internet Things},
month = jan,
articleno = {6},
numpages = {24},
}

@INPROCEEDINGS{aydogan2026impacts,
  author={Aydoğan, Didem and Chraiti, Mohaned and Tokgöz, Korkut Kaan},
  booktitle={2026 IEEE 37th International Symposium on Personal, Indoor and Mobile Radio Communications (PIMRC)}, 
  title={Impacts of Carrier Frequency and Numerology on Phase Noise Limited D-Band OFDM-ISAC Sensing}, 
  year={2026},
  volume={},
  number={},
  pages={1-6},
  doi={10.1109/PIMRC69400.2026.11711958}}

\end{document}